\documentclass[a4paper]{article}

\usepackage[utf8]{inputenc}
\usepackage[english]{babel}
\usepackage{amsmath} 
\usepackage{graphicx}
\usepackage{fancyhdr}
\usepackage{setspace}
\usepackage{lineno}	 
\usepackage{lastpage}
\usepackage[margin=1in]{geometry}
\usepackage{caption}  
\usepackage{amsmath,amssymb}
\usepackage{multirow}
\usepackage{makecell}
\usepackage{booktabs}
\usepackage[backend=biber,style=numeric-comp,sorting=none,autocite = superscript, maxnames=1, minnames=1]{biblatex}

{
	\fancyhead[L]{\myshorttitle}
	\fancyhead[R]{\small{\textit{\the\day.\the\month.\the\year}}}
	\fancyfoot[L]{\copyright \small{\textit{\myauthorshort}}}
	\fancyfoot[R]{Page \thepage\ of \pageref*{LastPage}}
	\cfoot{}
}

\fancypagestyle{firststyle}
{
	\fancyhead[L]{\copyright \small{\textit{\myauthorshort}}}
	\fancyhead[C]{}
	\fancyhead[R]{Page \thepage\ of \pageref*{LastPage}}
	\fancyfoot[L]{\footnotesize \rule{0.25\textwidth}{0.4pt} \newline Corresponding author: $^*$A. M. Schweidtmann, E-mail: a.schweidtmann@tudelft.nl}
	\fancyfoot[C]{}
	\fancyfoot[R]{}
}

\renewenvironment{abstract}{\noindent\textbf{Abstract:}}{}

\usepackage{authblk}
\usepackage{polski}
\usepackage{url}
\usepackage{algorithm}
\usepackage{algpseudocode}
\usepackage[version=4]{mhchem}
\usepackage{siunitx}
\usepackage{subcaption}
\usepackage{booktabs}
\usepackage{tabulary}
\usepackage{supertabular}
\usepackage{eurosym}
\usepackage{csquotes}
\algdef{SE}{Begin}{End}{\textbf{begin}}{\textbf{end}}
\DeclareSIUnit\year{a}
\newcolumntype{P}[1]{>{\centering\arraybackslash}p{#1}}
\newcolumntype{M}[1]{>{\centering\arraybackslash}m{#1}}
\newcolumntype{N}[1]{>{\raggedright\arraybackslash}m{#1}}

\renewbibmacro{finentry}{%
  \setunit{%
    \finentrypunct
    \renewcommand*{\finentry}{}%
    \par}%
  \usebibmacro{annotation}%
  \finentry
}

\DeclareBibliographyCategory{special}
\DeclareBibliographyCategory{outstanding}
\DeclareFieldFormat{labelnumber}{%
  \ifcategory{special}{#1\textbullet}{%
    \ifcategory{outstanding}{#1\textbullet\textbullet}{#1}}}

\newcommand{\mytitle}{Multi-agent systems for chemical engineering: A review and perspective}
\newcommand{\myshorttitle}{MASs for chemical engineering}
\newcommand{\myauthor}{Sophia Rupprecht $^{1}$, Qinghe Gao $^{1}$, Tanuj Karia $^{1}$, and Artur M. Schweidtmann $^{1,*}$} 
\newcommand{\myauthorshort}{ A. M. Schweidtmann}
\author{\myauthor}

\usepackage[colorlinks,linkcolor=blue,citecolor=blue,urlcolor=blue,pdftitle={\mytitle},pdfauthor={Artur M. Schweidtmann}]{hyperref} 
\usepackage{cleveref}

\begin{document}

\thispagestyle{firststyle}
	\begin{flushleft}\begin{large}\textbf{\mytitle}\end{large} \end{flushleft}
	\myauthor 
	
	\begin{flushleft}\begin{small}
			$^1$ Delft University of Technology, 
			Department of Chemical Engineering, 
			Van der Maasweg 9, 
			Delft 2629 HZ, 
			The Netherlands\\[0.15cm]
		\end{small}
	\end{flushleft}
\begin{abstract}
\noindent
Large language model (LLM)-based multi-agent systems (MASs) are a recent but rapidly evolving technology with the potential to transform chemical engineering by decomposing complex workflows into teams of collaborative agents with specialized knowledge and tools. This review surveys the state-of-the-art of MAS  within chemical engineering. While early studies demonstrate promising results, scientific challenges remain, including the design of tailored architectures, integration of heterogeneous data modalities, development of foundation models with domain-specific modalities, and strategies for ensuring transparency, safety, and environmental impact. As a young but fast-moving field, MASs offer exciting opportunities to rethink chemical engineering workflows.
\end{abstract}

\textbf{Keywords:} Process synthesis; large language models (LLMs); generative artificial intelligence (GenAI); agentic large language models; agentic artificial intelligence

\section{Introduction}
\label{2025MAFreview:Introduction}
The advent of large language models (LLMs), such as GPT‑4, has marked a technological leap, enabling advanced capabilities in language translation, multimodal data processing, mathematical reasoning, and code generation~\supercite{minaee2024large}. 
These advances have facilitated progress in various domains, including autonomous driving, chemistry, and materials science.
Likewise, there is growing interest in developing and applying LLMs in the chemical engineering domain~\supercite{Schweidtmann2024_GenAIinChemE, DecardiNelson2024_MultipleScales_GenAI4ChemE} and more specifically to process systems engineering~\supercite{DecardiNelson2024_GenAIPSE, Woo2025_GenAIAndLLM4PSE_SotAReview, Daoutidis2024_MLinPSE_ChallengesOpportunities}.

While LLMs have demonstrated strong capabilities in various domains, their application in chemical engineering remains limited due to multiple fundamental shortcomings.
LLMs are typically trained in an end-to-end fashion as black-box models, lacking transparency and interpretability.
Thereby, LLMs often fail to solve domain-specific, multi-step problems that demand physical reasoning and contextual integration. 
Chemical engineering workflows are inherently complex, requiring a systems-level perspective that spans multiple scales. 
These workflows rely heavily on specialized simulation tools and external data sources, which standalone LLMs cannot reliably replicate or replace, yet. 
Without access to relevant tools and trusted data sources, even the most advanced LLMs remain fundamentally constrained in their capabilities. 
Additionally, LLMs are prone to hallucinations arising from biases and gaps in their training data -- producing confident yet incorrect outputs. 
Such behavior leads to significant safety risks. 
Critically, general-purpose LLMs do not possess an intrinsic understanding of physical laws, such as thermodynamic constraints or mass and energy balances, frequently resulting in outputs that are physically invalid~\supercite{Venkatasubramanian2025_FromLLM2LargeKnowledgeModels, Du2025_PotentialAndChallengesOfLLMAgentSystemsInChemicalProcessSimulation_AutomatedModeling2IntelligentDesign}.

LLM-based multi-agent systems (MASs) offer a promising methodology to address these challenges. 
MASs subdivide complex tasks into smaller sub-tasks and assign roles to distinct agents that can interact with each other~\supercite{Wu2023_AutoGen, Liang2023_MultiAgentDebate}. 
Therein, single agents are usually LLMs that are equipped with external knowledge in the form of executable tools, access to databases, and literature~\supercite{minaee2024large}.
MASs have already shown success in other domains~\supercite{Tran2025_MultiAgentCollaboration_Survey, He2025_LLMBasedMultiAgentSystemsForSoftwareEngineering_Review}. 
Notable examples of MASs are in chemistry~\supercite{Ghafarollahi2024_ProtAgents, Boiko2023_AutonomousChemicalRes, Bran2024_ChemCrow}, drug discovery~\supercite{Fehlis2025_MultiAgentApproach4DrugDiscovery}, material design~\supercite{Tian2025_MultiAgentFramework4AcceleratedMetamaterialDesign}, scientific discovery~\supercite{Gottweis2025_AICoScientist, Lu2024_AIScientist, Yamada2025_AIScientist_v2}, medicine \supercite{Tang2023_MedAgents} and code generation~\supercite{Ishibashi2024_SelfOrganizedAgents_UltraLargeScaleCodeGenerationAndOptimization}.
Recent production-ready frameworks such as Clint and CodeRabbit further showcase the use of multi-agent collaboration in software engineering tasks, including autonomous development, code review, and project management.
Recently, several review and perspective articles have specifically focused on the potential of MAS in subfields of chemical engineering~\supercite{Du2025_PotentialAndChallengesOfLLMAgentSystemsInChemicalProcessSimulation_AutomatedModeling2IntelligentDesign, Boskabadi2025_IndustrialAgenticAIGenerativeModelingComplexSystems, Rothfarb2025_MultiAgentLLMFrameworks4OptimizingWatewaterTreatmentOperation}.
However, a comprehensive multi-agent framework for chemical engineering has not been proposed or developed yet.

In this article, we review the developments of MASs within the overarching chemical engineering domain and provide an outlook for further development and applications.
In Section~\ref{2025MAFreview:state_of_the_art}, we review the recent studies applying tailored, LLM-supported MASs demonstrating the potential of MAS for chemical engineering applications.
Furthermore, in Section~\ref{2025MAFreview:perspective}, we discuss the fundamental challenges that must be addressed to enable cross-scale integration in chemical engineering workflows with MASs.
Finally, we conclude the paper in Section~\ref{2025MAFreview:Conclusions}.

\section{State of the art}\label{2025MAFreview:state_of_the_art}
In this section, we first introduce the core concepts of MASs (Section~\ref{2025MAFreview:DefMAS}) and then review literature on applications of MAS in chemical engineering (Section~\ref{2025MAFreview:SotA_MASChemE}).

\subsection{Multi-agent systems}\label{2025MAFreview:DefMAS}
MAS are distributed systems composed of multiple autonomous agents that interact to achieve individual or collective goals~\supercite{minaee2024large}. 
Each agent can act independently by sensing the state of the environment to make decisions. 
While traditional supervised models learn static mappings, agents can dynamically reason over tasks, invoking tools or querying data as needed.
These capabilities facilitate iterative and reflective interaction in agents, thereby improving transparency and adaptability compared to standalone LLMs.
Hence, MASs provide a natural framework for streamlining chemical engineering workflows that require complex decision-making.
While MASs were proposed back in the 1990s for autonomous agents and distributed AI~\supercite{wooldridge2009introduction}, they have received significant attention recently due to the advent of LLMs.

For modern MASs, agents are typically LLMs capable of interacting with tools through structured application programming interfaces (APIs), providing inputs (e.g., JSON) and interpreting outputs.
Agents are also capable of querying structured information from relational databases (e.g., via SQL~\supercite{Hong2024_Text2SQL}) and knowledge graphs (e.g., via Cypher~\supercite{Ozsoy2024_Text2Cypher}), enabling access to structured domain knowledge.
For unstructured information—such as scientific literature, reports, or manuals—retrieval-augmented generation (RAG)~\supercite{Lewis2020_RAG} techniques are used to extract relevant content.
Large unstructured text corpora are segmented into smaller text chunks during preprocessing.
Each chunk is then transformed into a semantically rich vector representation, or embedding, and stored in a vector database.
RAG finally allows the retrieval of semantically similar text snippets during inference, given a user query.
Recent extensions adapt RAG for graph-based retrieval, i.e., Graph RAG,~\supercite{Edge2024_GraphRAG, Han2025_GraphRAG, Hu2024_GRAG} by transforming unstructured text into knowledge graphs.
The knowledge graphs used for Graph RAG-based approaches offer the advantage of capturing top-level relations between pieces of information, compared to isolated text snippets in the case of standard RAG methods~\supercite{Edge2024_GraphRAG}.

To facilitate communication among various agents to develop MASs, several open-source libraries such as AutoGen~\supercite{Wu2023_AutoGen}, crewAI~\supercite{CrewAI} by Microsoft, and LangGraph~\supercite{LangGraph_Github} by LangChain have been proposed. There exist many more open-source MAS frameworks, such as CAMEL~\supercite{Li2023_CAMEL} and MetaGPT~\supercite{Hong2023_MetaGPT}.
LangGraph~\supercite{LangGraph_Github} is widely used for scientific applications and supports flexible conversation via graph structures.
In LangGraph's graph structure, nodes represent agents, and edges represent communication.
LangGraph~\supercite{LangGraph_Github} is also capable of differentiating between (a) fixed (unconditional) and (b) dynamic (conditional) edges.
These frameworks can incorporate human agents, enabling expert oversight and intervention, enhancing reliability, interpretability, and control - crucial for safety-critical domains like chemical engineering.

\begin{figure}[h]
    \centering
    \includegraphics[width=\textwidth]{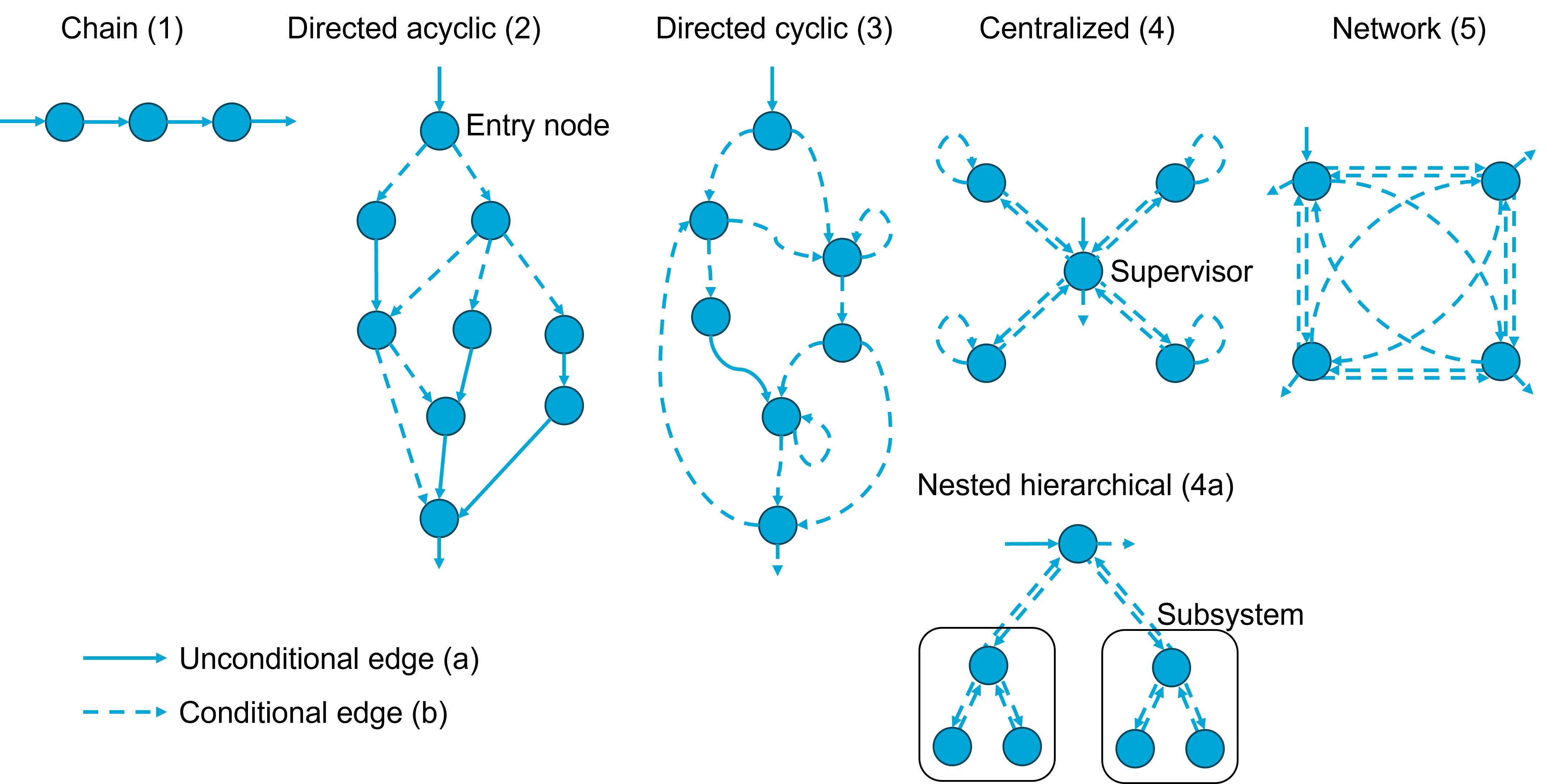}
    \caption{Architecture options for multi-agent systems. Multi-agent systems can be modeled as graphs~\supercite{LangGraph_Github} with agents as nodes and edges as conversation routes. Continuous edges (a) imply that there is a predetermined next agent, independent of the contribution of the previous agent, i.e., unconditional. Dashed edges (b) indicate that the next agent is dynamically inferred during inference depending on the state of the conversation, i.e., conditionally. The chain architecture (1) is a fixed sequence of agents connected with conditional edges. A directed acyclic conversation graph (2) contains conditional edges. A directed cyclic conversation graph (3) includes loops where substeps of the conversation can be repeated. In a centralized architecture (4), there is a single supervisor agent managing multiple subordinate agents. A nested hierarchical structure (4a) includes a supervisor managing multiple sub-systems, which can consist of single agents or independent, customized architectures (1-5). The network architecture (5) puts each agent in direct contact with every other agent independently.}
    \label{fig:MAS_architectures}
\end{figure}

Figure~\ref{fig:MAS_architectures} illustrates five basic architecture types of MASs~\supercite{LangGraph_Github, Tran2025_MultiAgentCollaboration_Survey, He2025_LLMBasedMultiAgentSystemsForSoftwareEngineering_Review, Han2024_MultiAgentSystems_ChallengesOpenProblems}:
(1) Chain: a linear sequence of agents using fixed edges only; 
(2) Directed acyclic graph (DAG): incorporates dynamic edges to enable branching paths; 
(3) Directed cyclic graph (DCG): allows iterative workflows by including cycles in the graph; 
(4) Centralized: a supervisor agent coordinates specialized sub-agents; 
(4a) Nested hierarchy: an extension of the centralized model where the supervisor manages modular sub-systems, each with its own internal structure; 
(5) Network: fully connected agents that communicate freely and can independently terminate or respond to tasks.
The choice of architecture depends on the nature of the application, including task complexity, communication needs, and coordination overhead. 
Each architecture offers trade-offs in terms of scalability, interpretability, and adaptability.

\subsection{Multi-agent systems in chemical engineering}\label{2025MAFreview:SotA_MASChemE}

In Table~\ref{tab:MAS_PSE}, we summarize recent studies showcasing the application of MASs for chemical engineering tasks~\supercite{Lee2024_PromptEngineeringLLMbasedProcessImprovement, Vyas2024_AutoControl_Agentic, Gowaikar2024_Text2PID, Sakhinana2024_PEOA, Pajak2025_MultiAgentLLMs4AutomatingSustainableOperationalDecisionMaking, Srinivas2025_AutoChemSchematic_AI, Zeng2025_LLMGuidedProcessOptimizationWithMultiAgentApproach, Srinivas2024_AcceleratingManufacturingScaleUp_AgenticWebNavigationAndRAG4ProcessEngineeringSchematicsDesign}.
\textcite{Lee2024_PromptEngineeringLLMbasedProcessImprovement} propose a MAS that suggests process design improvements based on existing PFDs. 
\textcite{Pajak2025_MultiAgentLLMs4AutomatingSustainableOperationalDecisionMaking} use a MAS to balance economic and environmental objectives in gas–oil separation. 
MASs have also been employed for general process problem-solving~\supercite{Sakhinana2024_PEOA}, autonomous industrial control~\supercite{Vyas2024_AutoControl_Agentic}, and generating PFDs or P\&IDs from natural language descriptions~\supercite{Gowaikar2024_Text2PID, Srinivas2025_AutoChemSchematic_AI}. 
Others use agents to identify operational constraints in process optimization~\supercite{Zeng2025_LLMGuidedProcessOptimizationWithMultiAgentApproach} or to build structured knowledge graphs from multimodal data for schematic synthesis via GraphRAG~\supercite{Srinivas2024_AcceleratingManufacturingScaleUp_AgenticWebNavigationAndRAG4ProcessEngineeringSchematicsDesign}. Now, we examine the publications along five aspects: (1) architecture of the MAS, (2) tool and (3) database integration, (4) modalities of data used, and (5) human integration.

\begin{table}[h]
\centering
\caption{Overview of multi-agent systems (MAS) in the chemical engineering domain. The existing literature is categorized by architecture type (chain, DAG (directed acyclic graph), DCG (directed cyclic graph), centralized), task addressed, tool and database integration (selection of examples), human integration, and framework used for implementation.}
\resizebox{\textwidth}{!}{%
\begin{tabular}{llcllcl}
\toprule
Literature & Task & MAS architecture & Tools (examples) & Database integration & Human integration & Framework \\
\midrule
\textcite{Lee2024_PromptEngineeringLLMbasedProcessImprovement} 
    & \makecell[l]{Process\\improvement} 
    & Chain 
    & \makecell[l]{ChemicalAid,\\Literature data extraction}
    & -
    & - 
    & OpenAI API \\
\addlinespace
\textcite{Vyas2024_AutoControl_Agentic} 
    & \makecell[l]{Industrial\\control} 
    & DCG 
    & \makecell[l]{Arduino temperature \\control lab} 
    & -
    & - 
    & CrewAI \\
\addlinespace
\textcite{Gowaikar2024_Text2PID} 
    & \makecell[l]{P\&ID generation\\from text} 
    & DCG 
    & \makecell[l]{MS Visio API} 
    & -
    & \checkmark 
    & n/a \\
\addlinespace
\textcite{Sakhinana2024_PEOA} 
    & \makecell[l]{General operations\\assistance} 
    & Centralized 
    & \makecell[l]{Stack Overflow,\\Wolfram Alpha} 
    & \makecell[l]{Graph RAG (Code \\ generation)}
    & - 
    & n/a \\
\addlinespace
\textcite{Pajak2025_MultiAgentLLMs4AutomatingSustainableOperationalDecisionMaking} 
    & \makecell[l]{Operational\\decision-making} 
    & DCG 
    & \makecell[l]{Pyomo,\\Aspen HSYS simulation} 
    & \makecell[l]{RAG (Economic reports, \\ COP28)}
    & - 
    & LangGraph \\
\addlinespace
\textcite{Srinivas2025_AutoChemSchematic_AI} 
    & \makecell[l]{PFD and P\&ID\\generation} 
    & DCG 
    & -
    & \makecell[l]{RAG (memory) \&\\ Graph RAG (process \\descriptions)}
    & \checkmark 
    & n/a \\
\addlinespace
\textcite{Zeng2025_LLMGuidedProcessOptimizationWithMultiAgentApproach} 
    & \makecell[l]{Process\\optimization} 
    & DCG
    & IDAES simulation
    & -
    & - 
    & AutoGen \\
\addlinespace
\textcite{Srinivas2024_AcceleratingManufacturingScaleUp_AgenticWebNavigationAndRAG4ProcessEngineeringSchematicsDesign} 
    & \makecell[l]{PFD and P\&ID\\generation} 
    & DAG 
    & \makecell[l]{Web search,\\Multimodal information extraction} 
    & \makecell[l]{Graph RAG (process \\descriptions)}
    & \checkmark
    & n/a \\
\bottomrule
\end{tabular}
}
\label{tab:MAS_PSE}
\end{table}

\paragraph{Architecture}\label{2025MAFreview:SotA_Architecture}
The DCG (3) is the most common architecture type of MAS in chemical engineering~\supercite{Vyas2024_AutoControl_Agentic, Gowaikar2024_Text2PID, Pajak2025_MultiAgentLLMs4AutomatingSustainableOperationalDecisionMaking, Srinivas2025_AutoChemSchematic_AI, Zeng2025_LLMGuidedProcessOptimizationWithMultiAgentApproach} (Table~\ref{tab:MAS_PSE}).
These systems typically mirror the structure of the engineering workflows they support. 
DCGs offer a balance between rigidity, which enforces task structure, and flexibility, which allows iteration and adaptation to edge cases.
This balance is critical in chemical engineering, where workflows must accommodate both structured procedures and case-specific adaptations.
By contrast, chain (1)~\supercite{Lee2024_PromptEngineeringLLMbasedProcessImprovement} and DAG (2) architectures enforce strictly linear or branching workflows without feedback loops. 
While these designs simplify agent interactions, they limit the system’s ability to handle iterative refinement or adapt to edge cases. 
This rigidity may constrain performance on more complex or ill-defined tasks.
Additionally, the centralized architecture (4), as used by \textcite{Sakhinana2024_PEOA}, delegates decision-making to a central coordinator agent, which favors flexibility over structured flow and aligns with tasks that lack a predefined engineering sequence.
Notably, network architectures (4a) and (5) have not yet been explored in MAS applications within chemical engineering.

\paragraph{Tool integration}
All reviewed MAS applications in chemical engineering rely on tool integration~\supercite{Lee2024_PromptEngineeringLLMbasedProcessImprovement, Vyas2024_AutoControl_Agentic, Gowaikar2024_Text2PID, Sakhinana2024_PEOA, Pajak2025_MultiAgentLLMs4AutomatingSustainableOperationalDecisionMaking, Srinivas2025_AutoChemSchematic_AI, Zeng2025_LLMGuidedProcessOptimizationWithMultiAgentApproach}, highlighting the central role of tool integration in enabling MAS functionality within chemical engineering
Previous work chose different ways to interface with the tools listed in Table~\ref{tab:MAS_PSE}.
For example, \textcite{Pajak2025_MultiAgentLLMs4AutomatingSustainableOperationalDecisionMaking} and \textcite{Zeng2025_LLMGuidedProcessOptimizationWithMultiAgentApproach} interact with simulation environments through pre-implemented interfaces.
\textcite{Sakhinana2024_PEOA}, on the other hand, connect to external APIs such as Stack Overflow and Wolfram Alpha.

\paragraph{Database integration}
Previous work on MAS in chemical engineering primarily interacts with vector databases~\supercite{Pajak2025_MultiAgentLLMs4AutomatingSustainableOperationalDecisionMaking, Sakhinana2024_PEOA, Srinivas2025_AutoChemSchematic_AI, Srinivas2024_AcceleratingManufacturingScaleUp_AgenticWebNavigationAndRAG4ProcessEngineeringSchematicsDesign}.
Naive RAG is used to extract relevant information from domain-specific documents~\supercite{Pajak2025_MultiAgentLLMs4AutomatingSustainableOperationalDecisionMaking}. 
Graph RAG database retrieval is applied for advanced code generation~\supercite{Sakhinana2024_PEOA} and retrieval of process flow and instrumentation descriptions~\supercite{Srinivas2025_AutoChemSchematic_AI, Srinivas2024_AcceleratingManufacturingScaleUp_AgenticWebNavigationAndRAG4ProcessEngineeringSchematicsDesign}.

\paragraph{Modalities}\label{2025MAFreview:review_modalities}
Chemical engineering spans heterogeneous data types: sensor signals, molecular representations (e.g., SMILES, graphs), process flowsheets, P\&IDs, and simulation files~\supercite{schweidtmann2021machine}. 
These modalities are typically preprocessed before being incorporated into the MAS or individual LLMs. 
Textual information (natural language instructions or contextual information) is often inserted directly into LLM prompts.
To combine other forms of modalities, such as tables, images, and code, a knowledge graph is employed~\supercite{Srinivas2025_AutoChemSchematic_AI, Sakhinana2024_PEOA, Srinivas2024_AcceleratingManufacturingScaleUp_AgenticWebNavigationAndRAG4ProcessEngineeringSchematicsDesign, Ovalle2025_GRAPSE}, thereby enabling the use of GraphRAG.
Numerical values - such as simulation results~\supercite{Pajak2025_MultiAgentLLMs4AutomatingSustainableOperationalDecisionMaking, Zeng2025_LLMGuidedProcessOptimizationWithMultiAgentApproach} or sensor data~\supercite{Vyas2024_AutoControl_Agentic} - are typically embedded into prompts. 
Lastly, processes are typically represented as graphs, which are used either as an alternative to text (e.g., D-SFILES~\supercite{Lee2024_PromptEngineeringLLMbasedProcessImprovement}) or as structured XML inputs~\supercite{Gowaikar2024_Text2PID, Alimin2025_talkingLikePID}. 

\paragraph{Human integration}
Most MASs in Chemical Engineering reviewed so far operate autonomously, and they do not integrate human oversight~\supercite{Lee2024_PromptEngineeringLLMbasedProcessImprovement, Vyas2024_AutoControl_Agentic, Sakhinana2024_PEOA, Pajak2025_MultiAgentLLMs4AutomatingSustainableOperationalDecisionMaking, Zeng2025_LLMGuidedProcessOptimizationWithMultiAgentApproach}.
Two studies embed human interaction into the MAS architecture~\supercite{Gowaikar2024_Text2PID, Srinivas2025_AutoChemSchematic_AI}:
\textcite{Gowaikar2024_Text2PID} employ a human assessor to validate new additions to a P\&ID visually through MS Visio.
\textcite{Srinivas2025_AutoChemSchematic_AI} use a human-in-the-loop to provide feedback on suggestions made during the solution process of the MAS.

Most works reviewed so far focus on solving a specific task in chemical engineering.
A notable exception is the work of ~\textcite{Sakhinana2024_PEOA}, which addresses general question answering in chemical engineering.
However, it is unclear how domain-specific tools and databases can be integrated into MASs for chemical engineering to enable multi-scale integration.
In the next section, we outline our perspective on developing a general MAS for chemical engineering.

\section{Perspective}\label{2025MAFreview:perspective}
We envision MASs to become an important tool for chemical engineering (Section~\ref{2025MAFreview:vision}). 
After proposing our vision, we discuss integral parts to achieve this vision, including MAS architecture (Section~\ref{2025MAFreview:architecture}), tool (Section~\ref{2025review:perspective_tools}) and database integration (Section~\ref{2025review:prespective_databases}), and chemical engineering foundation models with domain-specific modalities (Section~\ref{2025review:data_modalities}).
In Section~\ref{2025review:challenges}, we critically reflect on challenges regarding the transparency and reliability of such systems.

\subsection{Our vision}\label{2025MAFreview:vision}
We envision MAS in chemical engineering as interconnected, human-centric collaborators that integrate across different scales, with access to domain-specific tools, databases, and modalities to drive intelligent and transparent decision-making.

\begin{figure}[h]
    \centering
    \includegraphics[width=\textwidth]{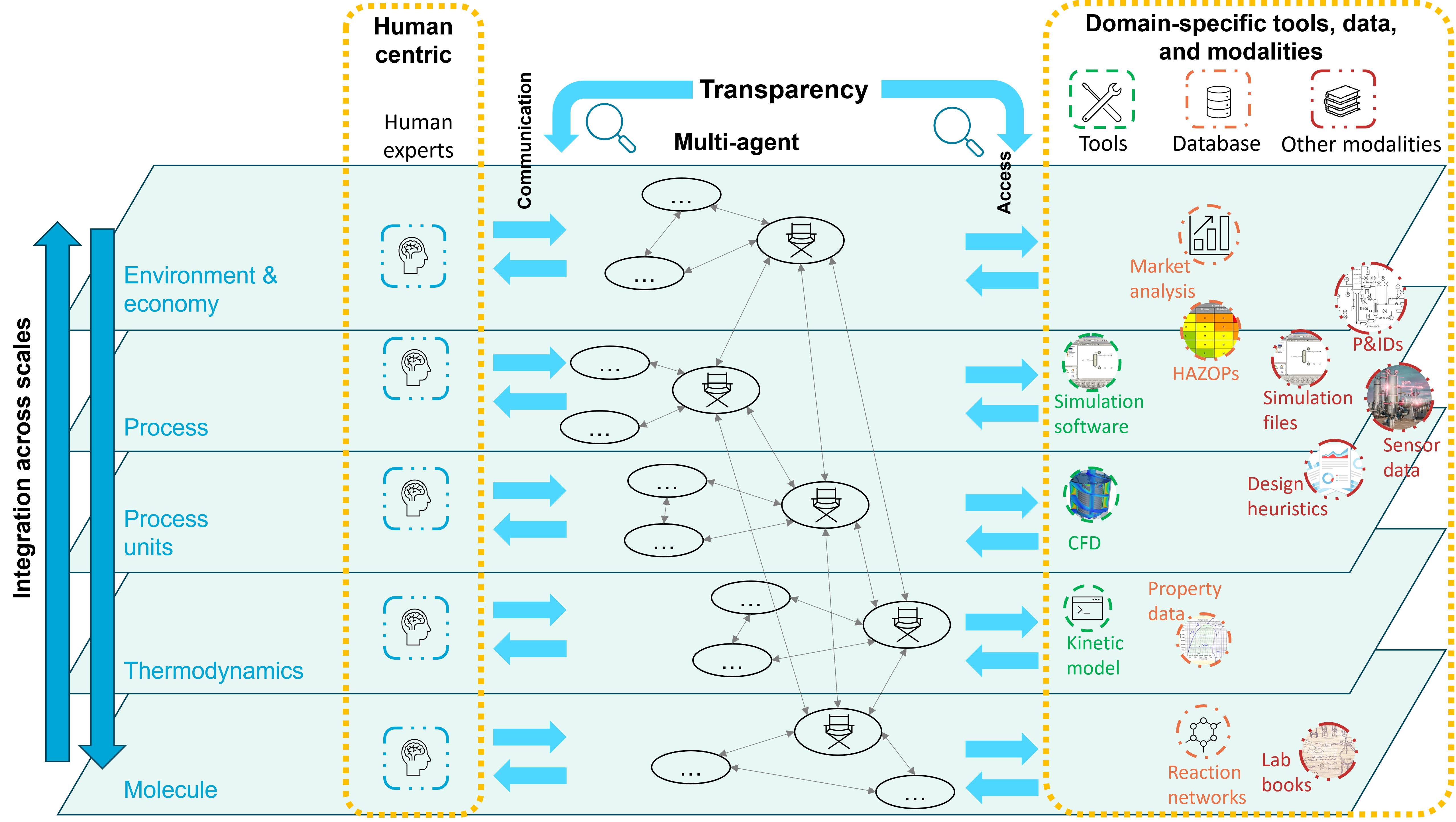}
    \caption{Multi-agent systems in chemical engineering as interconnected, human-centric collaborators that integrate across different scales, with access to domain-specific tools, databases, and modalities to drive intelligent and transparent decision-making.}
    \label{fig:MAS_overview}
\end{figure}

Chemical engineering challenges span multiple scales from molecular to plant-wide operations and global supply chains. 
As illustrated in Figure~\ref{fig:MAS_overview}, MASs offer a natural way to bridge these layers, with agents specializing in specific tasks while coordinating toward shared objectives. 
This architecture aligns closely with how chemical engineering work is already organized: in teams of experts using specialized tools and information. 
For example, agents equipped with thermodynamic and kinetic models can explore reaction pathways and materials, accelerating discovery at the molecular level. 
At the process level, operational agents dynamically run optimization routines, balancing yield, energy efficiency, and safety. 
Across plant and supply chain scales, agents negotiate schedules, logistics, and inventories in real time, while sustainability-focused agents continuously evaluate environmental and economic performance. 

We envision a future where every engineer manages a team of intelligent agents. 
This paradigm extends existing workflows and empowers engineers in a scalable, intuitive way. 
At all times, human engineers remain central to this ecosystem. 
Rather than replacing human expertise, multi-agent systems are designed to support and augment it. 
This requires agents to communicate in ways that are transparent, interpretable, and aligned with the users, e.g., through natural language, engineering diagrams, modeling code, and experimental procedures.

Realizing this vision depends on agents being well integrated into the chemical engineering domain. 
This includes interoperability with modeling and simulation tools, access to validated databases, and the ability to interpret diverse data modalities.
The goal is not to build generic end-to-end black-box AI systems, but domain-specialized teams of agents that understand and respect the intricacies of their subdomain within chemical engineering.
The following subsections outline the key building blocks and developments that are required to realize our vision.

\subsection{Architecture}\label{2025MAFreview:architecture}
The architecture of a MAS is a key design decision as it substantially influences the conversation workflows, adaptability, and overall performance~\supercite{Tran2025_MultiAgentCollaboration_Survey}.
Previous research has shown that architectural choices are a major contributing factor to the success of MASs in other domains~\supercite{Ishibashi2024_SelfOrganizedAgents_UltraLargeScaleCodeGenerationAndOptimization}.
For example, \textcite{Ishibashi2024_SelfOrganizedAgents_UltraLargeScaleCodeGenerationAndOptimization} proposes a flexible, hierarchical, dynamic, and scalable architecture for code generation and optimization, which yields competitive performance on a domain-specific benchmark.
Therefore, identifying suitable MAS architectures for chemical engineering is an important yet open research challenge.

Previous MAS architectures in chemical engineering have mirrored the existing engineering workflows (see Section~\ref{2025MAFreview:SotA_Architecture}).
While this approach facilitates adoption and provides a natural mapping between human and agent roles, it may also inherit the limitations of traditional process engineering methods. 
For example, hierarchical decomposition approaches, which are common in process design, can lead to suboptimal outcomes as early-stage decisions can negatively affect downstream design choices. 
MASs that strictly follow the hierarchical decomposition could suffer from the same shortcomings.
Instead, we envision a modular architecture that enables agents to coordinate flexibly while maintaining transparency and accountability. 
The modularity also ensures that the system can be easily extended towards additional tasks without re-designing the existing capabilities.

\subsection{Tool integration}\label{2025review:perspective_tools}
Tool integration is critical for MASs in chemical engineering, a domain reliant on specialized software across scales, i.e., from DFT at the molecular level to process simulators like Aspen Plus and gPROMS. 
Agents must interface with these tools via standardized endpoints (e.g., Python APIs), yet most commercial software lacks such access, posing a key barrier~\supercite{Du2025_PotentialAndChallengesOfLLMAgentSystemsInChemicalProcessSimulation_AutomatedModeling2IntelligentDesign}.
Therefore, developing open interfaces and data standards is essential. 
We envision MASs with tool-aware agents: some wrap specific software, while others orchestrate tools for higher-level reasoning.
Integration must be built into MAS architecture from the start to ensure scalability, interpretability, and reuse.
Future work should examine agent–tool interaction limits, architectural patterns, and strategies for managing tool complexity transparently for both agents and users.
Moreover, LLMs might require fine-tuning on our domain modeling languages to take full advantage of modeling environments beyond mere tool calls~\supercite{rupprecht2025text2model}.

\subsection{Database integration}\label{2025review:prespective_databases}
Access to structured, validated data is essential for MASs in chemical engineering, supporting accurate decision-making, reducing hallucinations, and enabling transparent reasoning. 
However, early MAS efforts have not yet considered database integration, increasing reliance on internal memory or LLM inference, which is prone to errors and unverifiable outputs.
We envision MASs interfacing with both open-access and commercial databases across scales: molecular (e.g., QM9, PubChem, reactions (e.g., ORD), thermodynamic (e.g., DIPPR, DDB), kinetic (e.g., NIST), and process-level (e.g., historian databases).
Future work should prioritize developing agents that access these databases via structured APIs or standardized interfaces. 
Embedding such capabilities early in MAS design will improve modularity, reasoning accuracy, and alignment with engineering practice.

\subsection{Chemical engineering foundation models with domain-specific modalities}\label{2025review:data_modalities}
For MASs to effectively support chemical engineering workflows, the underlying AI must process the field’s diverse data modalities (c.f. Section~\ref{2025MAFreview:review_modalities}).
AI cannot assist reliably if it cannot interpret the types of information engineers use daily.

A common approach to facilitate multimodal integration of previous approaches~\supercite{Ovalle2025_GRAPSE, Srinivas2025_AutoChemSchematic_AI} is to convert modalities to text, e.g., by using vision-language models to extract figure descriptions. 
Vision language models have been applied in chemical engineering, e.g., for the interpretation of scanning electron microscopy images~\supercite{Alshehri2023_MDL_ScientificImageInterpretation}. 
However, general-purpose LLMs have so far not shown sufficient capability to reliably handle various domain-specific modalities in a chemical engineering context (c.f., failure of visual language models in chemistry~\supercite{Alampara2024_MaCBench}). 
While allowing the use of LLMs out of the box, text representations have disadvantages when it comes to complex chemical engineering data. 

\begin{figure}[t]
    \centering
    \includegraphics[width=\textwidth]{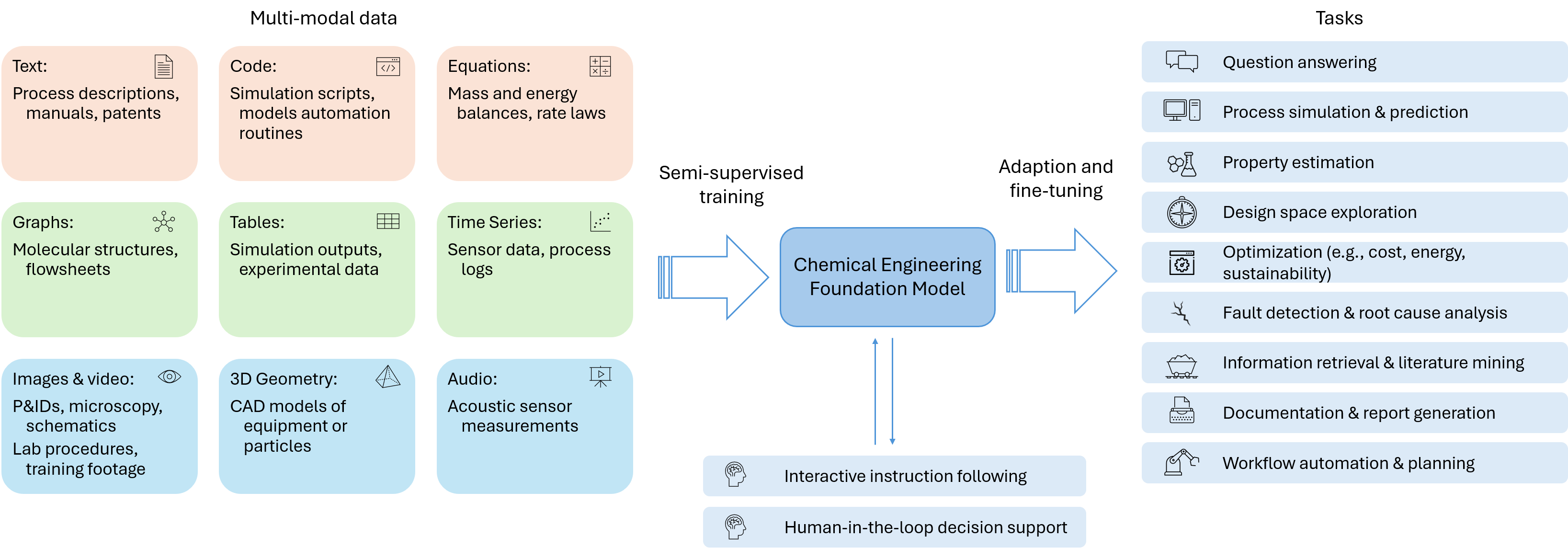}
    \caption{Modalities that need to be integrated into a chemical engineering foundation mode. On the left are common modalities and corresponding examples in the chemical engineering domain. A foundation model can be trained in a semi-supervised way. The foundation model can then be customized towards more specific downstream tasks.}
    \label{fig:MAS_modalities}
\end{figure}

We propose developing foundation models trained on chemical engineering–specific data (c.f. Figure~\ref{fig:MAS_modalities}). 
These models should incorporate multimodal learning, leveraging both labeled and unlabeled datasets through semi-supervised routines. 
In particular, we envision that graphs will play a key role as a data modality for chemical engineering~\supercite{gao2024deep,balhorn2025graph}, building on recent advances in computer science~\supercite{Ma2024_ToG2, Perozzi2024_LetYourGraphDoTheTalking, Jin2023_LLMsOnGraphs_Survey, Pan2023_UnifyingLLMsAndKGs, Luo2023_RoG}. 
Ontology-backed knowledge graphs can further reduce hallucinations through formal constraints and rule-based validation.
To get the engineering data in a machine-readable format, digitization efforts may be necessary~\supercite{Theisen2023_DigitizationOfChemicalProcessFlowDiagramsUsingDeepCNNs}. 
Such domain-specific foundation models could then be used as drop-ins for individual agents in MASs with the potential for fine-tuning in low-data regimes.
For industry use, compact models (e.g., small language models) or federated learning approaches may be necessary.

\subsection{Transparency, safety, and responsible development}\label{2025review:challenges}
MASs must meet the highest standards for transparency, reliability, and accountability, given the safety-critical nature of chemical processes. 
General-purpose LLMs, while powerful, are prone to hallucinations and lack grounding in physical laws, making them unsuitable as fully autonomous decision-makers in this domain.
Rather than aiming for full automation, MASs should be designed as symbiotic human-AI collaborations.
This includes static human-in-the-loop checkpoints and dynamic user consultation during inference - design principles that foster trust, ensure accountability, and align with emerging regulations like the EU AI Act, which mandates human oversight for high-risk AI applications.

MAS architectures must also be interpretable and evaluable. 
While recent open-source benchmark datasets provide a starting point~\supercite{Cemri2025_WhyDoMultiAgentLLMSystemsFail, Groetschla2025_AgentNets_CoordinationAndCollaborativeReasoningInMAS_Benchmark}, chemical engineering workflows require domain-specific performance criteria and failure taxonomies. Broader adoption will depend on standardized, task-specific evaluation strategies beyond isolated case studies~\supercite{Lee2024_PromptEngineeringLLMbasedProcessImprovement, Pajak2025_MultiAgentLLMs4AutomatingSustainableOperationalDecisionMaking}.
Finally, the development and deployment of LLM-based MASs must consider their computational and environmental costs. 
Training and operating large models consume significant energy, and these impacts should be assessed using life cycle analysis frameworks to ensure AI workflows support the broader sustainability goals of chemical engineering.

\section{Conclusions}
\label{2025MAFreview:Conclusions}
MASs are a very recent development in chemical engineering, with only a handful of pioneering studies published to date. 
Yet, they hold great promise to transform the entire field by essentially providing every engineer with a capable team of specialized agents. 
We surveyed MASs in chemical engineering, highlighting key developments in architecture, tool and database integration, addressing knowledge integration across modalities, and human-in-the-loop interaction. 
Despite early promise, the development is only starting, and scientific challenges remain, ranging from designing suitable architectures and integrating tools and data to developing foundation models with domain-specific modalities. 
In addition, issues of transparency, safety, human oversight, and environmental impact must be addressed.
As a young but fast-moving field, MASs offer exciting opportunities to rethink workflows in chemical engineering. 
\section{Acknowledgement}
This research is supported by Shell Global Solutions International B.V. and the Netherlands Organisation for Scientific Research (NWO) under the Veni grant, for which we express sincere gratitude.

\section{ Declaration of generative AI and AI-assisted technologies in the writing process}
During the preparation of this work, the author(s) used ChatGPT 4o in order to correct the grammar and polish the sentences. After using this tool/service, the author(s) reviewed and edited the content as needed and take(s) full responsibility for the content of the published article.
\printbibliography

\end{document}